\documentclass[draftcls, onecolumn, journal, 12pt, letterpaper]{IEEEtran}

\usepackage[dvips]{graphicx}
\graphicspath{{../eps/}}
\DeclareGraphicsExtensions{.eps}

\usepackage{multirow}
\usepackage{amsmath}
\usepackage{amsfonts}
\usepackage{amssymb}
\usepackage{longtable}

\usepackage{ifthen}
\newboolean{ONECOLUMN}
\setboolean{ONECOLUMN}{true}
\usepackage{verbatim}
\ifCLASSOPTIONcompsoc\usepackage[tight,normalsize,sf,SF]{subfigure}\else\usepackage[tight,footnotesize]{subfigure}\fi

\newenvironment{definition}[1][Definition]{\begin{trivlist}\item[\hskip \labelsep {\bfseries #1}]}{\end{trivlist}}

\usepackage[ruled, vlined]{algorithm2e}
\usepackage{hyperref}

\begin{document}
\title{Application of DAC Codeword Spectrum: Expansion Factor}
\author{Yong~Fang,~\IEEEmembership{Member,~IEEE}
\thanks{This research was supported by National Science Foundation of China (NSFC) (grant nos. 61001100, 61077009, and 60975007) and Provincial Science Foundation of Shaanxi, China (grant no. 2010K06-15).}
\thanks{The author is with the College of Information Engineering, Northwest A\&F University, Shaanxi Yangling 712100, China (email: yfang79@gmail.com; homepage: \url{http://www.wavesharp.com/fangyong/}).}%
}
\markboth{IEEE Transactions on Information Theory (Submission)}{Fang}
\maketitle

\begin{abstract}
Distributed Arithmetic Coding (DAC) proves to be an effective implementation of Slepian-Wolf Coding (SWC), especially for short data blocks. To study the property of DAC codewords, the author has proposed the concept of DAC codeword spectrum\footnotemark. For equiprobable binary sources, the problem was formatted as solving a system of functional equations. Then, to calculate DAC codeword spectrum in general cases, three approximation methods have been proposed. In this paper, the author makes use of DAC codeword spectrum as a tool to answer an important question: how many (including proper and wrong) paths will be created during the DAC decoding, if no path is pruned? The author introduces the concept of another kind of DAC codeword spectrum, i.e. \textbf{time spectrum}, while the originally-proposed DAC codeword spectrum is called \textbf{path spectrum} from now on. To measure how fast the number of decoding paths increases, the author introduces the concept of \textbf{expansion factor} which is defined as the ratio of path numbers between two consecutive decoding stages. The author reveals the relation between expansion factor and path/time spectrum, and proves that the number of decoding paths of any DAC codeword increases exponentially as the decoding proceeds. Specifically, when symbols `0' and `1' are mapped onto intervals $[0, q)$ and $[1-q, 1)$, where $0.5<q<1$, the author proves that expansion factor converges to $2q$ as the decoding proceeds.
\end{abstract}

\footnotetext[1]{In the previous papers on this topic, the author uses the terminology ``codeword distribution.'' To avoid such an ugly statement ``distribution of distributed ...,'' the author will use ``codeword spectrum'' to replace ``codeword distribution'' from now on.}


\section{Introduction}\label{sec:intro}
\IEEEPARstart{A}{rithmetic} Coding (AC) \cite{RissanenIBMJRD76} and its fast implementation Quasi AC (QAC) \cite{HowardITC92} have widely been used for data compression due to its entropy-approaching performance. To deal with noisy transmission, the AC can be extended in two ways to implement Joint Source-Channel Coding (JSCC): one is to introduce forbidden intervals corresponding to forbidden symbols \cite{BoydTC97, PettijohnTC01}, e.g. Error-Correcting AC (ECAC), which has been used for image and video transmission \cite{GuionnetTIP03, GrangettoCL03, GrangettoTIP06, GrangettoTIP07}; the other is to insert markers into the sequence of source symbols at fixed positions \cite{SodagarICASSP00}. Recently, to deal with Slepian-Wolf Coding (SWC) \cite{SlepianIT73}, the AC has also been extended in two ways: one is to introduce overlapped intervals corresponding to ambiguous symbols, e.g. Distributed AC (DAC) \cite{GrangettoCL07, GrangettoTSP09} and Overlapped QAC (OQAC) \cite{ArtigasICIP07}; the other is to puncture some bits of AC bitstream, e.g. Punctured QAC (PQAC) \cite{MalinowskiPCS09}. There are also some variants of the DAC, e.g. Time-Shared DAC (TS-DAC) \cite{GrangettoMMSP07} for symmetric SWC, rate-compatible DAC \cite{GrangettoCL08}, decoder-driven adaptive DAC \cite{GrangettoICIP08} for online estimation of source statistics, etc. Most recently, DAC implementation of Distributed Joint Source-Channel Coding (DJSCC) has also appeared \cite{GrangettoICIP10}.

Let $\boldsymbol{x}$ be the source and $\boldsymbol{y}$ decoder Side Information (SI). It is straightforward to know that the performance of the AC is possible to approach to source entropy $H(\boldsymbol{x})$. However, to the best of the author's knowledge, no analysis on the performance of the ECAC and the DAC is found in the literature up to now. For the ECAC, we have no idea whether the rate can approach to the limit $H(\boldsymbol{x})/C$, where $C$ is channel capacity. For the DAC, nobody knows whether the rate can approach to the limit $H(\boldsymbol{x}|\boldsymbol{y})$ \cite{SlepianIT73}. For the DAC-based DJSCC, it remains an open question whether the rate can approach to the limit $H(\boldsymbol{x}|\boldsymbol{y})/C$.

This paper is devoted to the performance analysis on the DAC. In the author's opinion, to answer the question whether the rate of the DAC can approach to the limit $H(\boldsymbol{x}|\boldsymbol{y})$, the prerequisites include two folds. First, one needs to know how many paths will be created as the DAC decoding proceeds. Second, one should know the Probability Density Function (PDF) of the Hamming distances between those decoding paths and the source.

In \cite{FangSPL09}, the author introduces the concept of codeword spectrum which is a function defined over interval $[0, 1)$. For DAC codeword spectrum of equiprobable binary sources along proper decoding paths, the problem is formatted as solving a system of functional equations including four constraints \cite{FangSPL09}. Then, three approximation methods are proposed in \cite{FangTIT10} for calculating DAC codeword spectrum, i.e. numeric approximation, polynomial approximation, and Gaussian approximation. Though the concept of DAC codeword spectrum seems wonderful, it finds no usage in practice up to now.

In this paper, by using DAC codeword spectrum as a tool, the author answers an important question: how many (proper and wrong) paths will be created as the DAC decoding proceeds? This is the first application of DAC codeword spectrum up to now. Through this work, the author expects to find more applications of DAC codeword spectrum in the future.

This paper is arranged as follows. Section \ref{sec:principle} briefly introduces the principle of DAC codec. Section \ref{sec:preliminaries} introduces the concepts that will be used in the following analyses, e.g. \textbf{path spectrum}, \textbf{time spectrum}, \textbf{population}, \textbf{expansion factor}, etc., and reveals the relations between expansion factor and path/time spectrum. Section \ref{sec:time} researches the evolution and numeric calculation of time spectrum. Section \ref{sec:results} reports experimental and theoretical results of expansion factor. Finally, Section \ref{sec:conclusion} concludes this paper.

\section{Principle of Distributed Arithmetic Coding}\label{sec:principle}

\subsection{Encoding}\label{sec:principle_enc}
Consider an infinite-length, stationary, and equiprobable binary source $\boldsymbol{x} = \{x_i\}_{i=1}^{\infty}$. Let $\boldsymbol{y} = \{y_i\}_{i=1}^{\infty}$ be decoder SI, where $\Pr(x_i \neq y_i) = p$. The DAC encoder \cite{GrangettoCL07, GrangettoTSP09} iteratively maps source symbols `0' and `1' onto intervals $[0, q)$ and $[(1-q), 1)$, where $q = 2^{-\alpha}$. We call $\alpha$ overlapping factor, which satisfies 
\begin{equation}
	\frac{H(\boldsymbol{x}|\boldsymbol{y})}{H(\boldsymbol{x})} = H(\boldsymbol{x}|\boldsymbol{y}) \leq \alpha \leq 1. 
\end{equation}
The resulting codeword of $\boldsymbol{x}$ is denoted by $C_\alpha(\boldsymbol{x})$. The DAC encoding process is in fact a transform that converts source $\boldsymbol{x}$ into codeword $C_\alpha(\boldsymbol{x})$. We denote the rate of $C_\alpha(\boldsymbol{x})$ by $R_{\alpha}(\boldsymbol{x})$. It is easy to obtain
\begin{equation}
	R_{\alpha}(\boldsymbol{x}) = \alpha H(\boldsymbol{x}) = \alpha \geq H(\boldsymbol{x}|\boldsymbol{y}).
\end{equation}

\subsection{Decoding}\label{sec:principle_dec}
The DAC decoder works in a symbol-driven mode. Because $(1-q) < q$ when $q \in (0.5, 1)$, intervals $[0, q)$ and $[(1-q), 1)$ are partially overlapped. Though this overlapping leads to a larger final interval and hence a shorter codeword, it also causes an ambiguity during the decoding as a cost. To describe the DAC decoding process, a ternary symbol set $\{0, \mathcal{A}, 1\}$ is defined, where $\mathcal{A}$ represents the ambiguous symbol. Once symbol $\mathcal{A}$ is met, the decoder will perform a branching: two candidate paths are created, corresponding to two alternative symbols `0' and `1'. Therefore, as the decoding proceeds, more and more paths will be created. Undoubtedly, among them, there is only one proper path corresponding to source $\boldsymbol{x}$. We denote the $j$-th path as $\tilde{\boldsymbol{x}}_j = \{\tilde{x}_{ji}\}_{i=1}^{\infty}$, where $\tilde{x}_{ji}$ is the $i$-th symbol along path $\tilde{\boldsymbol{x}}_j$.

For path $\tilde{\boldsymbol{x}}_j$, when decoding symbol $\tilde{x}_{ji}$, the state of a $B$-bit DAC decoder is described by parameter set $(l_{ji}, h_{ji}, c_{ji})$, where $l_{ji}$, $h_{ji}$, and $c_{ji}$ are $B$-bit integers. $l_{ji}$ and $h_{ji}$ are the lower and upper bounds of the range at time $i$. $c_{ji}$ is the $B$-bit codeword in the buffer at time $i$. Obviously,
\begin{equation}
	0 \leq l_{ji} \leq c_{ji} \leq h_{ji} \leq (2^B - 1).
\end{equation}
Let
\begin{equation}
	u_{ji} = \frac{c_{ji} - l_{ji}}{h_{ji} - l_{ji} + 1}.
\end{equation}
Then
\begin{equation}
	\setlength{\nulldelimiterspace}{0pt}
	\tilde{x}_{ji} = \left\{
		\begin{IEEEeqnarraybox} [\relax] [c] {l's}
			0, &$0 \leq u_{ji} < (1-q)$ \\
			\mathcal{A}, &$(1-q) \leq u_{ji} < q$\\
			1, &$q \leq u_{ji} < 1$%
		\end{IEEEeqnarraybox}.
	\right.
\end{equation}

If $\tilde{x}_{ij} = \mathcal{A}$, then two candidate paths are created, corresponding to symbols `0' and `1', respectively. For each path, its metric is updated according to SI $\boldsymbol{y}$ and its corresponding sub-interval is selected for next iteration. To maintain linear complexity, each time a symbol is decoded, the decoder makes use of the $M$-algorithm to retain at most $M$ paths with the best partial metric, and prune others \cite{GrangettoCL07, GrangettoTSP09}. Finally, after all source symbols are decoded, the path with the best overall metric is output as the estimate of $\boldsymbol{x}$.

\section{Preliminaries}\label{sec:preliminaries}

\begin{figure*}
\centering
\ifCLASSOPTIONtwocolumn
	\includegraphics[width=.7\linewidth]{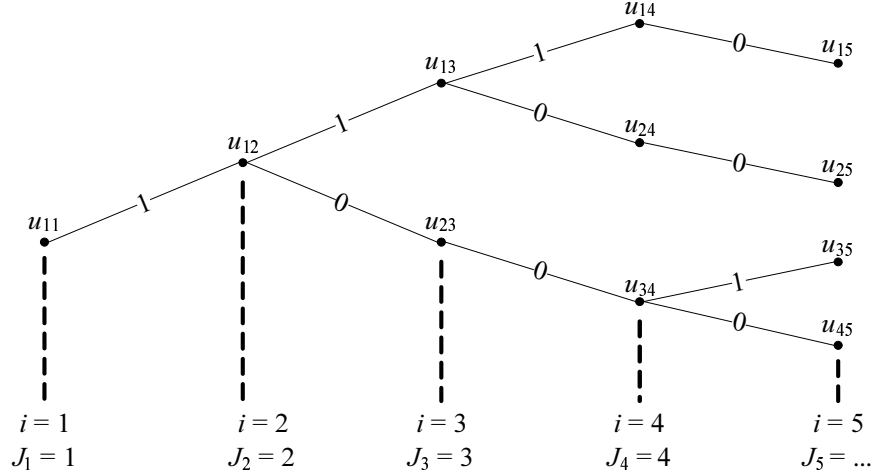}
\else
	\includegraphics[width=.7\linewidth]{decoding.eps}
\fi
\caption{Illustration of the concepts of path, path spectrum, population, time spectrum, and expansion factor. In this example, there are four decoding paths: $\tilde{\boldsymbol{x}}_1 = ``1110"$, $\tilde{\boldsymbol{x}}_2 = ``1100"$, $\tilde{\boldsymbol{x}}_3 = ``1001"$, and $\tilde{\boldsymbol{x}}_4 = ``1000"$, each of which corresponds to its path spectrum, e.g., the path spectrum along path $\tilde{\boldsymbol{x}}_4$ is the PDF of $\boldsymbol{u}_{4*} = (u_{11}, u_{12}, u_{23}, u_{34}, u_{45})$. The population increases as the decoding proceeds, e.g., there are three paths after time $i=3$, so $J_3 = 3$ (initially, $J_0 \equiv 1$). Each decoding time corresponds to a time spectrum, e.g., the time spectrum at time $i=3$ is the PDF of $\boldsymbol{u}_{*3} = \{u_{13}, u_{23}\}$. According to the definition of expansion factor, we have $\gamma_1 = J_1/J_0 = 1$, $\gamma_2 = 2$, $\gamma_3 = 3/2$, etc.}
\label{fig:decoding}
\end{figure*}

\subsection{Definitions}\label{sec:preliminaries_def}
With the help of Fig. \ref{fig:decoding}, we give the definitions of path spectrum, population, time spectrum, and expansion factor in turn as follows.

\begin{definition}\label{def:path_spectrum}{\textit{\textbf{Path Spectrum}:}}
When decoding $C_\alpha(\boldsymbol{x})$ along path $\tilde{\boldsymbol{x}}_j$, the PDF of $\boldsymbol{u}_{j*} = \{u_{ji}\}_{i=1}^{\infty}$ is called the path spectrum of $C_\alpha(\boldsymbol{x})$ along path $\tilde{\boldsymbol{x}}_j$. 
\end{definition}

For example, in Fig. \ref{fig:decoding}, there are four decoding paths, each of which corresponds to its path spectrum. Specially, we are interested in the path spectrum along the proper decoding path $\boldsymbol{x}$, which is denoted by $f(u)$, where $u \in [0, 1)$. According to \cite{FangSPL09}, $f(u)$ should satisfy the following constraints
\begin{equation}
	\setlength{\nulldelimiterspace}{0pt}
	\begin{IEEEeqnarraybox}[\relax][c]{l's}
		\int_0^1 {f(u)du} = 1\\
		f(u) = f(1-u)\\
		f(u) = f(u/q)/(2q), & $0 \leq u < (1-q)$ \\
		f(u) = \frac{f(\frac{u}{q}) + f(\frac{u-(1-q)}{q})}{2q}, & $(1-q) \leq u < q$ \\ 
		f(u) = \frac{f(\frac{u-(1-q)}{q})}{2q}, & $q \leq u < 1$
	\end{IEEEeqnarraybox}.
\end{equation}
As for the calculation of $f(u)$, three approximation methods have been proposed in \cite{FangTIT10}.

\begin{definition}\label{def:population}{\textit{\textbf{Population}:}}
The number of paths \textbf{after} decoding the $i$-th symbol is called the population at time $i$, which is denoted by $J_i$.
\end{definition}

As there is only one path \textbf{before} decoding the first symbol, we have $J_0 \equiv 1$. As for the example of population, please refer to Fig. \ref{fig:decoding}.

\begin{definition}\label{def:breadth_spectrum}{\textit{\textbf{Time Spectrum}:}}
When decoding the $i$-th symbol, there are $J_{i-1}$ paths. We call the PDF of $\boldsymbol{u}_{*i} = \{u_{ji}\}_{j=1}^{J_{i-1}}$ as the time spectrum of $C_\alpha(\boldsymbol{x})$ at time $i$, which is denoted by $g_{i}(u)$. 
\end{definition}

Please refer to Fig. \ref{fig:decoding} for the example of time spectrum. As $g_i(u)$ is the PDF of $u$, the normalization property should hold, i.e.
\begin{equation}
	\int_0^1 {g_i(u)du} = 1.
\end{equation}
In addition, as we are investigating equiprobable binary sources, the symmetry property should also hold, i.e.
\begin{equation}
	g_i(u) = g_i(1-u).
\end{equation}
When decoding the first symbol, there is only one path, which is undoubtedly the proper path. Hence, from the statistical view, the time spectrum at time $i=1$ is equivalent to the path spectrum along the proper decoding path $\boldsymbol{x}$, i.e.
\begin{equation}
	g_1(u) = f(u).
\end{equation}

\begin{definition}\label{def:expansion_factor}{\textit{\textbf{Expansion Factor}:}}
We define the expansion factor at time $i$ as the ratio of the expectation of $J_{i}$ to that of $J_{i-1}$, which is denoted by $\gamma_i$, i.e. $\gamma_i = \frac{E(J_{i})}{E(J_{i-1})}$.
\end{definition}

Please refer to Fig. \ref{fig:decoding} for the example of expansion factor.

\subsection{Relations between Population, Expansion Factor, and Time Spectrum}\label{sec:preliminaries_relations}
When $u_{ji}$ falls into $[(1-q), q)$, two branches will be created, or in other word, one more path will be created. Therefore, if there are $J_{i-1}$ paths at time $(i-1)$, then from the statistical view, $J_{i-1}(\int_{1-q}^{q}{g_{i}(u)du})$ more paths will be created at time $i$ on average, i.e.
\begin{equation}
	E(J_{i}) = E(J_{i-1}) (1 + \int_{1-q}^{q}{g_{i}(u)du}).
\end{equation}
Therefore, the expansion factor at time $i$ is 
\begin{equation}
	\gamma_i = \frac{E(J_{i})}{E(J_{i-1})} = 1 + \int_{1-q}^{q}{g_{i}(u)du}.
\end{equation}
Especially, as $g_1(u) = f(u)$ and $J_0 \equiv 1$, we have
\begin{equation}
	\gamma_1 = E(J_1) = 1 + \int_{1-q}^{q}{f(u)du}.
\end{equation}
Then recursively, we have
\begin{equation}\label{eqn:population}
	E(J_{i}) = \prod_{i'=1}^{i}{\gamma_{i'}}.
\end{equation}

From the above analyses, we can see that time spectrum $g_i(u)$ is the key to answering all questions. Once we know $g_i(u)$, expansion factor $\gamma_i$ can be obtained and then population $E(J_i)$ can be deduced in turn.

\section{Time Spectrum}\label{sec:time}
\subsection{Evolution}\label{sec:time_evolution}
\begin{figure*}
\centering
\ifCLASSOPTIONtwocolumn
	\includegraphics[width=\linewidth]{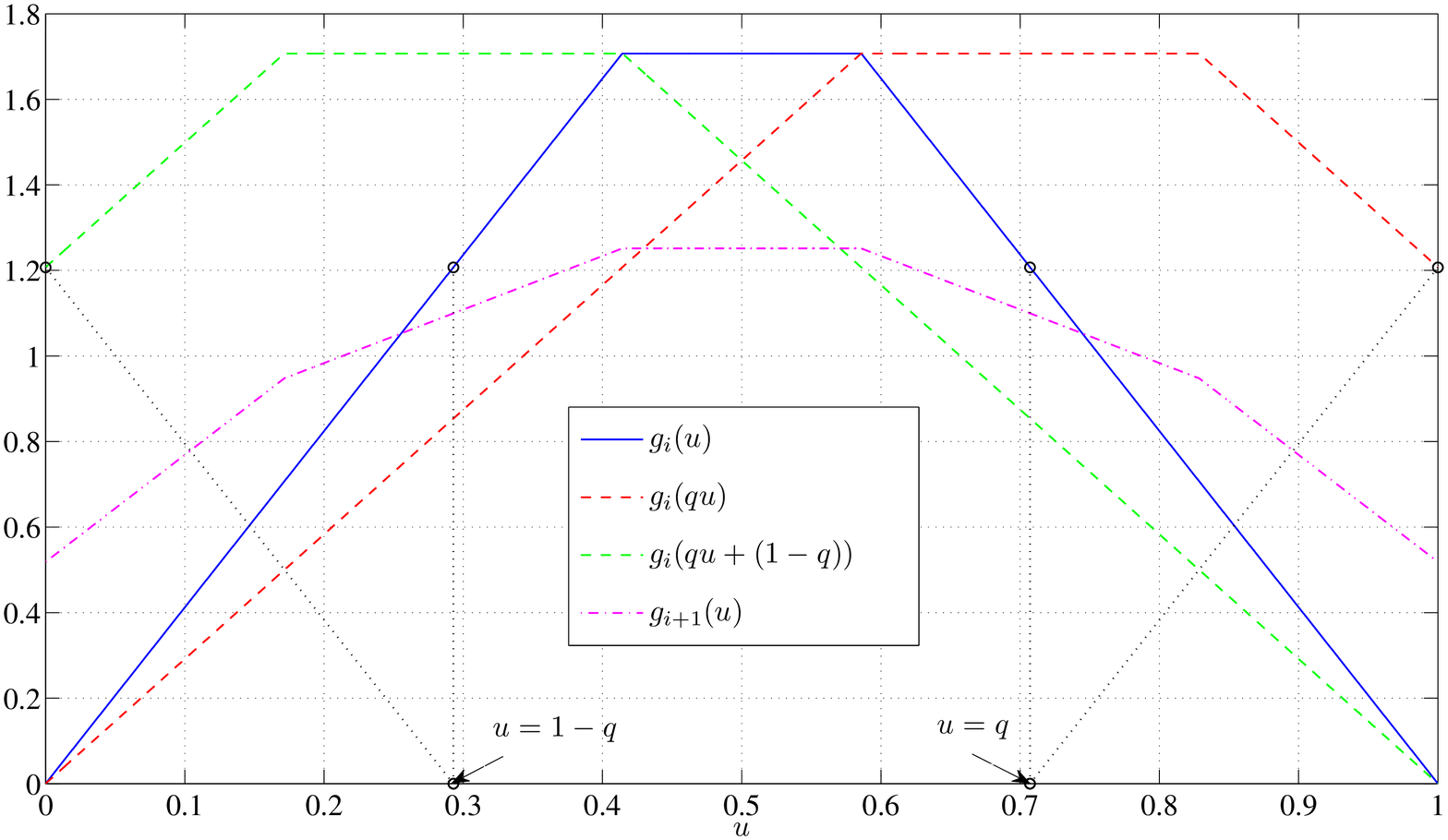}
\else
	\includegraphics[width=\linewidth]{evolution.eps}
\fi
\caption{Illustration of the evolution of time spectrum. In this example, $q = 1/\sqrt{2}$ and $i = 1$, so $g_i(u) = g_1(u) = f(u)$, where the closed form of $f(u)$ has been obtained in \cite{FangSPL09}. When $0 \leq u < q$, the 0-branch will be created and then interval $[0, q)$ will be mapped onto interval $[0, 1)$ at the next iteration (as shown by $g_i(qu)$). Similarly, when $(1-q) \leq u < 1$, the 1-branch will be created and then interval $[(1-q), 1)$ will be mapped onto interval $[0, 1)$ at the next iteration (as shown by $g_i(qu+(1-q))$). Therefore, $g_{i+1}(u)$ should be the normalized sum of $g_i(qu)$ and $g_i(qu+(1-q))$.}
\label{fig:evolution}
\end{figure*}

With the help of Fig. \ref{fig:evolution}, we illustrate how the time spectrum evolutes as the decoding proceeds. Let $g_i(u)$ be the time spectrum at time $i$. If $0 \leq u < q$, then the 0-branch will be created and interval $[0, q)$ at time $i$ will be mapped onto interval $[0, 1)$ at time $(i+1)$. It means that the part of $g_i(u)$ over interval $0 \leq u < q$ will be mapped onto $g_i(qu)$ over interval $0 \leq u < 1$ at the next iteration [Fig. \ref{fig:evolution}]. Similarly, if $(1-q) \leq u < 1$, then the 1-branch will be created and interval $[(1-q), 1)$ at time $i$ will be mapped onto interval $[0, 1)$ at time $(i+1)$. Meanwhile, the part of $g_i(u)$ over interval $(1-q) \leq u < 1$ will be mapped onto $g_i(qu+(1-q))$ over interval $0 \leq u < 1$ at the next iteration [Fig. \ref{fig:evolution}]. Finally, the time spectrum at time $(i+1)$ should be the sum of $g_i(qu)$ and $g_i(qu+(1-q))$ [Fig. \ref{fig:evolution}], i.e.
\begin{equation}
	g_{i+1}(u) = \beta_i(g_{i}(qu) + g_{i}(qu + (1-q))),
\end{equation}
where $\beta_i$ is introduced to make sure $\int_{0}^{1}{g_{i+1}(u)du} = 1$. It is easy to obtain
\begin{equation}
	\beta_i = \frac{q}{1 + \int_{1-q}^q {g_{i}(u)du}} = q/\gamma_i.
\end{equation}
As $i$ approaches to the infinite, we have
\begin{equation}
	g_{\infty}(u) = \beta_{\infty}(g_{\infty}(qu) + g_{\infty}(qu + (1-q))), \quad \forall u \in [0, 1).
\end{equation}
Hence,
\begin{equation}
	g_{\infty}(u) \equiv 1, \quad \forall u \in [0, 1).
\end{equation}
It means: as the decoding proceeds, the time spectrum will converge to the uniform distribution. Meanwhile, we can also obtain $\beta_{\infty} = 1/2$. Finally
\begin{equation}
	\gamma_{\infty} = 1 + \int_{1-q}^{q}{g_{\infty}(u)du} = 2q = 2^{1-\alpha}.
\end{equation}

\subsection{Discussion}\label{sec:time_discussion}
Intuitively, $E(J_i)$ reflects the residual uncertainty of $\boldsymbol{x}$ given its DAC codeword $C_\alpha(\boldsymbol{x})$. Therefore, the conditional entropy of $\boldsymbol{x}$ given $C_\alpha(\boldsymbol{x})$ can be calculated by
\begin{equation}
	H(\boldsymbol{x}|C_\alpha(\boldsymbol{x})) = \lim_{i \rightarrow \infty}{\frac{\log_2{E(J_{i})}}{i}}.
\end{equation}
According to (\ref{eqn:population}), we have
\begin{eqnarray}
	\log_2{E(J_{i})} = \log_2{\prod_{i'=1}^{i}{\gamma_{i'}}} = \sum_{i'=1}^{i}{\log_2{\gamma_{i'}}}.
\end{eqnarray}
Thus,
\begin{equation}\label{eqn:uncertainty}
	H(\boldsymbol{x}|C_\alpha(\boldsymbol{x})) = \lim_{i \rightarrow \infty}{\frac{\sum_{i'=1}^{i}{\log_2{\gamma_{i'}}}}{i}} = 1-\alpha.
\end{equation}
It is obvious that
\begin{equation}
	H(\boldsymbol{x}|C_\alpha(\boldsymbol{x})) = H(\boldsymbol{x}) - I(\boldsymbol{x}; C_\alpha(\boldsymbol{x})),
\end{equation}
i.e.
\begin{equation}
	1-\alpha = 1 - I(\boldsymbol{x}; C_\alpha(\boldsymbol{x})).
\end{equation}
Thus, we obtain
\begin{equation}
	I(\boldsymbol{x}; C_\alpha(\boldsymbol{x})) = \alpha.
\end{equation}
Since $C_\alpha(\boldsymbol{x})$ is the codeword of $\boldsymbol{x}$, the mutual information between $C_\alpha(\boldsymbol{x})$ and $\boldsymbol{x}$ is just the partial information of $\boldsymbol{x}$ provided $C_\alpha(\boldsymbol{x})$. Recall that the rate of $C_\alpha(\boldsymbol{x})$ is $R_\alpha(\boldsymbol{x}) = \alpha$, so
\begin{equation}
	R_\alpha(\boldsymbol{x}) = I(\boldsymbol{x}; C_\alpha(\boldsymbol{x})).
\end{equation}
It means that the rate of a DAC codeword can reach the mutual information between it and the coded source, or in other word, any rate-$\alpha$ DAC codeword conveys $\alpha$ bits information of the coded source on average.

\subsection{Numeric Approximation}\label{sec:time_numeric}
As path spectrum $f(u)$, to find the closed form of time spectrum $g_i(u)$ is not an easy thing. Thus, inspired by the work in \cite{FangTIT10}, the author proposes a numeric method for calculating $g_i(u)$. This method is described in detail below.

\subsubsection{Discretization}
We divide the interval $[0, 1]$ into $N$ uniform cells. Let $\Delta = 1/N$. Then $g_i(u)$ can be approximated by $g_i(n\Delta)$, where $n \in \mathcal{I}_N = \{0,1,...,N\}$, for a large $N$.

\subsubsection{Initialization}
Before iteration, we set $g_1(n\Delta) = f(n\Delta)$, $\forall n \in \mathcal{I}_N$, where $f(n\Delta)$ can be obtained by the method given in \cite{FangTIT10}.

\subsubsection{Update}
Recursively, $g_{i+1}(n\Delta)$ can be obtained from $g_i(n\Delta)$ by (we omit coefficient $\beta_i$)
\begin{equation}
	g_{i+1}(n\Delta) = g_i(round(nq)\Delta) + g_i(round(nq + N(1-q))\Delta).
\end{equation}

\subsubsection{Normalization}
As $\int_{0}^{1}{g_{i+1}(u)du} = 1$, we have $\sum_{n=0}^{N}{g_{i+1}(n\Delta)\Delta} = 1$, i.e. 
\begin{equation}
	\sum_{n=0}^{N}{g_{i+1}(n\Delta)} = 1/\Delta = N. 
\end{equation}
Let $\sum_{n=0}^{N}{g_{i+1}(n\Delta)} = \Omega$, then $g_{i+1}(n\Delta)$ should be normalized as
\begin{equation}
	g_{i+1}(n\Delta) = \frac{N}{\Omega}g_{i+1}(n\Delta).
\end{equation}

\subsubsection{Expansion Factor}
Let $L = round(N(1-q))$ and $H = round(Nq)$, then the expansion factor at time $(i+1)$ can be calculated by
\begin{equation}\label{eqn:expansion_factor_numeric}
	\gamma_{i+1} = 1 + \frac{\sum_{n=L}^{H}{g_{i+1}(n\Delta)}}{N}.
\end{equation}

\section{Simulation Results}\label{sec:results}
\begin{figure*}
\centering
\ifCLASSOPTIONtwocolumn
	\includegraphics[width=\linewidth]{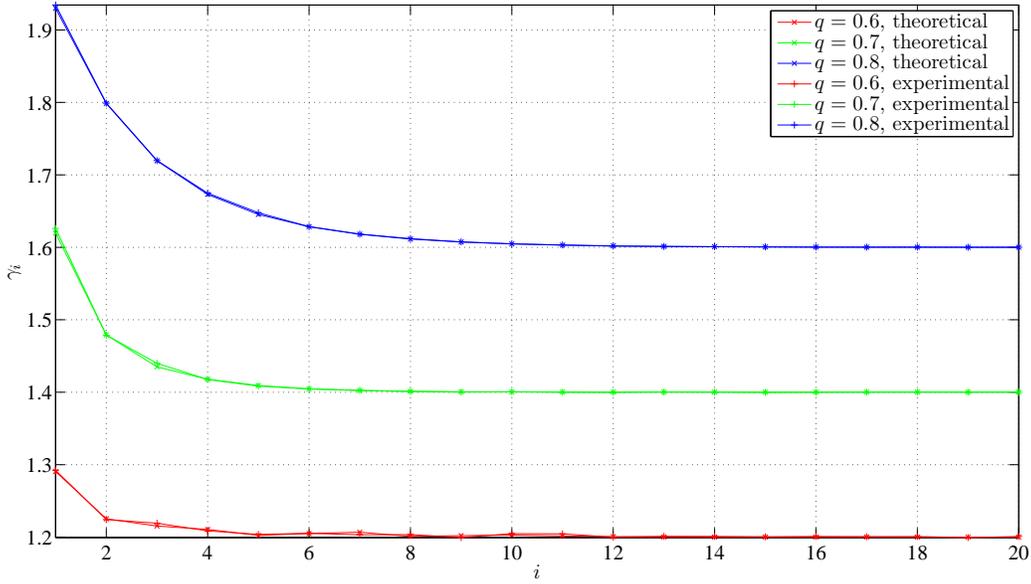}
\else
	\includegraphics[width=\linewidth]{expanding_factor.eps}
\fi
\caption{Theoretical and experimental results of expansion factor for $q=0.6$, 0.7, and 0.8. A 31-bit DAC codec is used for the experiments and the results are averaged over $10^4$ DAC codewords of various length-1024 equiprobable binary sequences. For the theoretical results, the number of cells is $N=10^5$. The software of theoretical results is available on the author's homepage.}
\label{fig:results}
\end{figure*}

Fig. \ref{fig:results} includes some theoretical and experimental results of expansion factor. For theoretical results, the author first calculates the path spectrum along the proper decoding path $f(u)$ through the numeric method given in \cite{FangTIT10}, where the number of cells is set to $N=10^5$. Then seeded with $f(u)$, the numeric method given in Section \ref{sec:time_numeric} is run to obtain $g_i(u)$, where the number of cells is also set to $N=10^5$. Finally, the expansion factor at time $i$ is obtained by (\ref{eqn:expansion_factor_numeric}).

For experimental results, a 31-bit DAC codec is used to encode $10^4$ various length-1024 equiprobable binary sequences. Then these codewords are decoded. The decoder first counts the number of length-$i$ paths (i.e. only $i$ symbols are decoded for each path), $J_i$, through full search. Then the expansion factor at time $i$ can be obtained by $\gamma_i = \frac{E(J_i)}{E(J_{i-1})}$, where $E(J_i)$ means the average of $J_i$ over $10^4$ DAC codewords. 

From Fig. \ref{fig:results}, the reader can find that the theoretical results coincide with the experimental results perfectly. Both theoretical and experimental curves converge to $2q$ rapidly, meaning that the above analyses are well verified.

\section{Conclusion}\label{sec:conclusion}
This paper researches an important problem: how many paths will be created as the DAC decoding proceeds? To answer this question, the author inctroduces the concepts of path spectrum, time spectrum, and expansion factor. The relations between time spectrum, path spectrum, and expansion factor are revealed. A numeric method to calculate time spectrum is proposed. The given experimental and theoretical results coincide with each other perfectly. In the future, the author will continue the work and research another important problem: how about the PDF of the Hamming distances between decoding paths and the source?




\begin{thebibliography}{20}

\bibitem{RissanenIBMJRD76}
	J. J. Rissanen,
	\newblock ``Generalized \textsc{K}raft inequality and arithmetic coding,''
	\newblock {\em {IBM} J. Research and Development}, vol. 20, no. 3, pp. 198--203, May 1976.

\bibitem{HowardITC92}
	P. G. Howard and J. S. Vitter,
	\newblock ``Practical implementations of arithmetic coding,''
	\newblock in: {\em Image and Text Compression}, pp. 85--112, Kluwer Academic, Norwell, Mass, USA, 1992.
	
\bibitem{BoydTC97}
	C. Boyd, J. G. Cleary, S. A. Irvine, I. Rinsma-Melchert, and I. H. Witten,
	\newblock ``Integrating error detection into arithmetic coding,''
	\newblock {\em{IEEE} Trans. Commun.}, vol. 45, no. 1, pp. 1--3, Jan. 1997.

\bibitem{PettijohnTC01}
	B. D. Pettijohn, M. W. Hoffman, and K. Sayood,
	\newblock ``Joint source/channel coding using arithmetic codes'',
	\newblock {\em{IEEE} Trans. Commun.}, vol. 49, no. 5, pp. 826--836, May 2001.

\bibitem{GuionnetTIP03}
	T. Guionnet and C. Guillemot,
	\newblock ``Soft decoding and synchronization of arithmetic codes: application to image transmission over noisy channels,''
	\newblock {\em {IEEE} Trans. Image Process.}, vol. 12, no. 12, pp. 1599--1609, Dec. 2003.

\bibitem{GrangettoCL03}
	M. Grangetto, E. Magli, and G. Olmo,
	\newblock ``Robust video transmission over error-prone channels via error correcting arithmetic codes,''
	\newblock {\em{IEEE} Commun. Lett.}, vol. 7, no. 12, pp. 596--598, Dec. 2003.

\bibitem{GrangettoTIP06}
	M. Grangetto, E. Magli, and G. Olmo,
	\newblock ``A syntax preserving error resilience tool for \textsc{JPEG}2000 based on error correcting arithmetic coding,''
	\newblock {\em{IEEE} Trans. Image Process.}, vol. 15, no. 4, pp. 807--818, Apr. 2006.

\bibitem{GrangettoTIP07}
	M. Grangetto, B. Scanavino, G. Olmo, and S. Bendetto,
	\newblock ``Iterative decoding of serially concatenated arithmetic and channel codes with \textsc{JPEG}2000 applications,''
	\newblock {\em{IEEE} Trans. Image Process.}, vol. 16, no. 6, pp. 1557--1567, Jun. 2007.

\bibitem{SodagarICASSP00}
	I. Sodagar, B. B. Chai, and J. Wus,
	\newblock ``A new error resilience technique for image compression using arithmetic coding,'' 
	\newblock in: {\em Proc. IEEE ICASSP}, pp. 2127--2130, Istanbul, Turkey, June 2000.

\bibitem{SlepianIT73}
	D.~Slepian and J. K.~Wolf,
	\newblock ``Noiseless coding of correlated information sources,''
	\newblock {\em {IEEE} Trans. Inf. Theory}, vol. 19, no. 4, pp. 471--480, July 1973.

\bibitem{GrangettoCL07}
	M.~Grangetto, E.~Magli, and G.~Olmo,
	\newblock ``Distributed arithmetic coding,''
	\newblock {\em {IEEE} Commun. Lett.}, vol. 11, no. 11, pp. 883--885, Nov. 2007.

\bibitem{GrangettoTSP09}
	M. Grangetto, E. Magli, and G. Olmo,
	\newblock ``Distributed arithmetic coding for the \textsc{S}lepian-\textsc{W}olf problem,''
	\newblock {\em{IEEE} Trans. Signal Process.}, vol. 57, no. 6, pp. 2245--2257, Jun. 2009. 

\bibitem{ArtigasICIP07}
	X.~Artigas, S.~Malinowski, C.~Guillemot, and L.~Torres,
	\newblock ``Overlapped quasi-arithmetic codes for distributed video coding,''
	\newblock in: {\em Proc. IEEE ICIP}, 2007, vol.~II, pp. 9--12.

\bibitem{MalinowskiPCS09}
	S.~Malinowski, X.~Artigas, C.~Guillemot, and L.~Torres,
	\newblock ``Distributed coding using punctured quasi-arithmetic codes for memory and memoryless sources,''
	\newblock {\em IEEE Trans. Signal Process.}, vol. 57, no. 10, pp. 4154--4158, Oct. 2009.

\bibitem{GrangettoMMSP07}
	M.~Grangetto, E.~Magli, and G.~Olmo,
	\newblock ``Symmetric distributed arithmetic coding of correlated sources,''
	\newblock in: {\em Proc. IEEE MMSP}, 2007, pp. 111--114.

\bibitem{GrangettoCL08}
	M.~Grangetto, E.~Magli, R.~Tron, and G.~Olmo,
	\newblock ``Rate-compatible distributed arithmetic coding,''
	\newblock {\em {IEEE} Commun. Lett.}, vol. 12, no. 8, pp. 575--577, Aug. 2008.

\bibitem{GrangettoICIP08}
	M.~Grangetto, E.~Magli, and G.~Olmo,
	\newblock ``Decoder-driven adaptive distributed arithmetic coding,''
	\newblock in: {\em Proc. IEEE ICIP}, 2008, pp. 1128--1131.
	
\bibitem{GrangettoICIP10}
	M. Grangetto, E. Magli, and G. Olmo,
	\newblock ``Distributed joint source-channel arithmetic coding,''
	\newblock {\em IEEE ICIP}, 2010, to be presented,	available online: \url{http://www.di.unito.it/~mgrange/}
	
\bibitem{FangSPL09}
	Y.~Fang,
	\newblock ``Distribution of distributed arithmetic codewords for equiprobable binary sources,''
	\newblock {\em {IEEE} Signal Process. Lett.}, vol. 16, no. 12, pp. 1079--1082, Dec. 2009.
	
\bibitem{FangTIT10}
	Y.~Fang,
	\newblock ``Approximation of DAC codeword distribution for equiprobable binary sources along proper decoding path,''
	\newblock {\em {IEEE} Trans. Inf. Theory}, submitted, available online: \url{http://arxiv.org/abs/1009.5257v1}.
	
	
	

\end{thebibliography}
\end{document}